\documentstyle[amsfonts,eqsecnum,aps]{revtex}

\begin{document}
\draft
\title{Confined Harmonically Interacting Spin-Polarized Fermions }
\author{F.~Brosens 
\thanks{Senior Research Associate of the FWO (Fons voor Wetenschappelijk Onderzoek-Vlaanderen).}%
and J. T. Devreese 
\thanks{Also at Universiteit Antwerpen (RUCA) and Technische Universiteit Eindhoven, Nederland.}%
}
\address{Departement Natuurkunde, Universiteit Antwerpen (UIA),\\
Universiteitsplein 1, B-2610 Antwerpen}
\author{L. F. Lemmens}
\address{Departement Natuurkunde, Universiteit Antwerpen (RUCA),\\
Groenenborgerlaan 171, B-2020 Antwerpen}
\date{}
\maketitle

\begin{abstract}
The thermodynamical properties are calculated for a three-dimensional model
of $N$ harmonically interacting spin-polarized fermions in a parabolic
potential well. The obtained dependences of the chemical potential and of
the internal energy on the complete range of the temperature and of the
number of particles turn out to obey a scaling law, similar to the scaling
from the continuum approximation for the density of states. The calculation
technique is based on our path integral approach of symmetrized density
matrices for identical particles in a parabolic confining well.
\end{abstract}


\section{Introduction}

In two preceding papers the present authors extended the method of
symmetrized density matrices to systems confined in a parabolic well\cite
{BDLPRE97a} (hereafter referred to as I) and used this method to obtain
expressions for the density and the pair correlation function \cite
{BDLPRE97b} (hereafter referred to as II). The evaluation of the internal
energy, the specific heat, the moment of inertia \cite{Stringari,BDLPRA97},
the density and the static response functions was performed for bosons,
inspired by the recently observed Bose-Einstein condensation\cite
{BEC1,BEC2,BEC3} and the theoretical work around this phenomenon using other
methods \cite{Grossman,Grossman2,Ketterle,Kirsten,Haugerud,CohenLee,Krauth}.
In both papers I and II the general expressions for most of the quantities
mentioned above are also given for Fermi-Dirac statistics. In the present
paper a method is presented to explicitly evaluate the thermodynamic
quantities for spin-polarized fermions. The model is a parabolic well
containing $N$ fermions all in the same spin state and interacting through a
harmonic two-body potential that may be either attractive or repulsive.

A quantum dot would be a physical system that could be described by such a
model, if also a magnetic field is taken into account to freeze away the
opposite spin states. When no magnetic field is taken into account two spin
states (spin up and spin down) should be present in the model. Recently\cite
{Butts} it was proposed to investigate also confined fermions in the same
experimental configuration used for the bose-alkali metals. The Thomas-Fermi
approximation\cite{Silvera} was used to study the spatial distribution of
these trapped fermion systems. In order to analyze these more physically
relevant models in the future, we first develop in the present communication
the required basic techniques for spin-polarized fermions. The model has
also some importance in itself, because it can be used to test new
approaches to Monte Carlo simulations of interacting fermions such as
many-body diffusion\cite{LBD,BDLssc,LBDPR}.

The paper is organized as follows. In section II, we collect the expressions
from I and II for the fermion case, and we show in section III how the
chemical potential, the free energy and the internal energy can be obtained
for a given number of fermions as a function of the temperature.
Subsequently the low temperature limit is considered, and the groundstate
energy is evaluated in section IV. In the last section a discussion and the
conclusions are given.

\section{Fermion oscillators}

In this section the basic formulas which have been derived in the
path-integral treatment of I and II are summarized and rewritten in such a
way that they are more appropriate for dealing with fermions, in particular
for the numerical treatment. Before doing so, it is instructive to point out
where the numerical accuracy problems are coming from. Having pinpointed
their origin, a method is proposed to accurately evaluate the relevant
thermodynamic quantities.

\subsection{Summary of previous results}

We consider a model of $N$ fermions with parallel spin in an harmonic
confinement potential, and with a quadratic interparticle interaction. The
one-body potential energy $V_1$ and the two-body potential energy $V_2$ of
the model system are given by 
\begin{equation}
V=V_1+V_2;\quad V_1=\frac{m\Omega ^2}2\sum_{j=1}^N\left. {\bf r}_j\right.
^2;\quad V_2=-\frac{m\omega ^2}4\sum_{j,l=1}^N\left( {\bf r}_j-{\bf r}%
_l\right) ^2.
\end{equation}
The two-body interaction is assumed to be repulsive; replacing $-\omega ^2$
by $\omega ^2$ in $V_2$ gives the case of attraction. In {\sl each dimension}
we found one degree of freedom (the center of mass) with frequency $\Omega ,$
and $N-1$ degrees of freedom with frequency $w$ given by 
\begin{equation}
w=\sqrt{\Omega ^2-N\omega ^2},
\end{equation}
which means that the frequency $\Omega $ of the center of mass is larger
than the frequencies $w$ of the degrees of freedom in the relative
coordinate system. Changing the sign of $\omega ^2$ allows to obtain the
case with $w$ larger than $\Omega .$

In our path-integral treatment presented in I, a recurrence relation was
obtained for the partition function ${\Bbb Z}_I\left( N\right) $
corresponding to the degrees of freedom with frequency $w$ in the
relative-coordinate system. Introducing 
\begin{equation}
b=e^{-\beta \hbar w}
\end{equation}
for brevity in the notations, we found that: 
\begin{equation}
{\Bbb Z}_I\left( N\right) =\frac 1N\sum_{m=0}^{N-1}\xi ^{N-m-1}\left( \frac{%
b^{\frac 12\left( N-m\right) }}{1-b^{N-m}}\right) ^3{\Bbb Z}_I\left(
m\right) .  \label{eq:ZIrecur}
\end{equation}
This recurrence relation applies for bosons ($\xi =+1$) and for fermions ($%
\xi =-1$). The subscript $I$ refers to identical particles, which can be
specified to be fermions (subscript $F$) or bosons (subscript $B$). The
total partition function $Z_I\left( N\right) $ only differs from ${\Bbb Z}%
_I\left( N\right) $ by a factor which accounts for the center-of-mass
contribution 
\begin{equation}
Z_I\left( N\right) =\left( \frac{\sinh \frac 12\beta w}{\sinh \frac 12\beta
\Omega }\right) ^3{\Bbb Z}_I\left( N\right) .
\end{equation}

\subsection{The ``sign''--problem and the canonical ensemble}

For three-dimensional (3D) fermions, the contribution (\ref{eq:ZIrecur}) to
the partition function from the relative degrees of freedom clearly
illustrates the kind of numerical inaccuracies which originate for the
fermion case $\xi =-1.$ If the partition functions for $1,$ $2,$ ..., $%
\left( N-1\right) $ particles are known, Cramer's rule can be used to
calculate the partition function for $N$ fermions. Factorizing the
denominators in the partition function (\ref{eq:ZIrecur}) by introducing the
quantities 
\begin{equation}
z_N=b^{-\frac 32N}{\Bbb Z}_F\left( N\right) \prod_{j=1}^N\left( 1-b^j\right)
^3,
\end{equation}
a careful analysis shows that $z_N$ are polynomials in $b$. Typical terms of
the expansion in powers of $b$ are summarized in table \ref{TablezN}. In
table \ref{Tablez10} the polynomial $z_{10}$ is given in full detail.

These expressions clearly illustrate that the recurrence relation (\ref
{eq:ZIrecur}) with its alternating signs is numerically not stable: the
leading terms are of order $b^M$ where $M$ increases drastically if the
number of fermions increases. Nevertheless, the expression for $z_{10}$,
e.g., is useful to check expressions for the partition function or derived
quantities on their accuracy. Because for the fermions, unlike the boson
case, solving the recurrence relations thus runs into severe numerical
problems, we will use the generating function technique for the actual
calculation of the free energy and the internal energy. To convince
ourselves that numerical inaccuracies have been avoided, the internal energy
of the model for up to 10 fermions has been calculated both ways, i.e. from
the recurrence relations and with the generating function technique, and the
results of both methods coincide. How the chemical potential and the
internal energy are calculated will be elaborated in the next section.

\section{The thermodynamic properties}

The generating function $\Xi _F\left( u\right) $ corresponding to the
partition functions ${\Bbb Z}_F\left( N\right) $ is defined in the standard
way as 
\begin{equation}
\Xi _F\left( u\right) =\sum_{N=0}^\infty u^N{\Bbb Z}_F\left( N\right) .
\label{eq:KsiTaylor}
\end{equation}
As shown in I, it is given by 
\begin{equation}
\Xi _F\left( u\right) =\exp \left( -\sum_{j=1}^\infty \frac 1j\frac{\left(
-ub^{\frac 32}\right) ^j}{\left( 1-b^j\right) ^3}\right) .
\label{eq:Ksicyclic}
\end{equation}
This means that in our model the internal degrees of freedom are represented
by a system of non-interacting oscillators with frequency $w.$ $\Xi _F\left(
u\right) $ is then {\sl formally} the grand-canonical partition function of
that subsystem. However, it is {\sl not} the grand-canonical partition
function of the {\sl full model} system with interaction for two reasons:
first one has to take the center-of-mass correction into account, and second
the eigenfrequency $w$ in the relative coordinate system depends on the
number of particles. But {\sl given} $w$ the full mechanism of the
generating functions is applicable in the relative coordinate system,
provided afterwards the necessary center-of-mass corrections are taken into
account.

The partition function ${\Bbb Z}_F\left( N\right) $ from the internal
degrees of freedom can be obtained by inverting the defining Taylor series (%
\ref{eq:KsiTaylor}) 
\begin{equation}
{\Bbb Z}_F\left( N\right) =\frac 1{2\pi i}\oint_c\frac{\Xi _F\left( z\right) 
}{z^{N+1}}dz
\end{equation}
where $C$ is a closed contour in the complex $z$ plane around the origin.
The generating function $\Xi _F\left( z\right) $ is unaccessible for
numerical purposes. However, considering a circular contour with radius $u$
one obtains 
\begin{equation}
{\Bbb Z}_F\left( N\right) =\frac 1{2\pi }\int_0^{2\pi }\frac{\Xi _F\left(
ue^{i\theta }\right) }{u^N}e^{-iN\theta }d\theta =\frac 1{2\pi }\int_0^{2\pi
}\exp \left[ \ln \Xi _F\left( ue^{i\theta }\right) -N\ln u\right]
e^{-iN\theta }d\theta .
\end{equation}
The extremum of $\left[ \ln \Xi _F\left( ue^{i\theta }\right) -N\ln u\right] 
$ on the real axis satisfies the condition $N=u\frac d{du}\ln \Xi _F\left(
u\right) .$ Using (\ref{eq:Ksicyclic}) this requirement becomes 
\begin{equation}
N=\sum_{\nu =0}^\infty n_\nu ,\quad n_\nu =\frac 12\frac{\left( \nu
+1\right) \left( \nu +2\right) }{1+e^{-\beta \left( \mu -\epsilon _\nu
\right) }},\quad \epsilon _\nu =\hbar w\left( \nu +\frac 32\right) ,
\label{eq:mufromN}
\end{equation}
which is precisely the result which one would obtain from the ``grand
canonical'' treatment with $u=e^{\beta \mu },$ taking into account the
degeneracy $\frac 12\left( \nu +1\right) \left( \nu +2\right) $ of the $\nu $%
th energy level. Factorizing out the steepest descent contribution $\Xi
_F\left( u\right) /u^N$ obtained this way, one finds 
\begin{eqnarray}
{\Bbb Z}_F\left( N\right) &=&\frac{\Xi _F\left( u\right) }{u^N}\int_0^\pi
\Psi \left( \theta \right) d\theta ,  \label{eq:Zintegral} \\
\Psi \left( \theta \right) &=&\frac 1\pi \frac{\Xi _F\left( ue^{i\theta
}\right) }{\Xi _F\left( u\right) }e^{-iN\theta },  \label{eq:Psi}
\end{eqnarray}
where $\Psi \left( \theta \right) $ is a real function, suitable for
numerical integration if $u=e^{\beta \mu }$ is determined.

The advantage of a procedure based on the generating function is that all
contributions to $\Xi _F\left( u\right) $ turn out to be positive, in
contrast to the direct determination of the partition function (\ref
{eq:ZIrecur}) which numerically involves\thinspace severe sign problems, as
argued in the previous section.

\subsection{The chemical potential}

The chemical potential has to be determined from the requirement (\ref
{eq:mufromN}). There are clearly two cases to be considered. For
sufficiently low temperature, $\mu $ will be larger than $\frac 32\hbar w,$
but at high temperature $\mu $ might be smaller than $\frac 32\hbar w$.

For the case $\mu >\frac 32\hbar w,$ the behavior of the denominator is
fundamentally different for the energy levels $\epsilon _\nu <\mu $ as
compared to those satisfying $\epsilon _\nu >\mu .$ Assume that $\mu $ will
be found in the interval $]\epsilon _{L-1},\epsilon _L]$ between two
consecutive levels 
\begin{equation}
\mu =\epsilon _L-\hbar w\alpha ;\text{\quad }\alpha \in \left[ 0,1\right[ ,
\label{eq:mu}
\end{equation}
and treat the levels $\epsilon _\nu \leq \mu $ separately from those with $%
\epsilon _\nu >\mu .$ We start the discussion under the assumption $L\geq 1$
and rewrite $N$ as: 
\[
N=\sum_{\nu =1}^L\frac 12\frac{\left( L-\nu +1\right) \left( L-\nu +2\right) 
}{1+b^{-\alpha +\nu }}+\frac 12\frac{\left( L+1\right) \left( L+2\right)
b^\alpha }{1+b^\alpha }+\sum_{\nu =1}^\infty \frac 12\frac{\left( L+\nu
+1\right) \left( L+\nu +2\right) b^{\alpha +\nu }}{1+b^{\alpha +\nu }}. 
\]
Let $N_L$ denote the (not necessarily integer) value which the right hand
side of this equation would take for $\alpha =0,$ i.e. if the chemical
potential would exactly be located at $\epsilon _L.$ One finds after some
algebra (provided $L>0$) 
\[
N_L=\frac 16\left( L+\frac 32\right) \left( L+2\right) \left( L+1\right)
+\left( 2L+3\right) \sum_{\nu =1}^L\frac{\nu b^\nu }{1+b^\nu }+\sum_{\nu
=L+1}^\infty \frac 12\frac{\left( L+\nu +1\right) \left( L+\nu +2\right)
b^\nu }{1+b^\nu }>\frac 16L^3. 
\]
The result $L<\left( 6N_L\right) ^{1/3}$ is important for developing a
sure-fire algorithm to find the chemical potential. If one starts from a
value $L$ which is the largest integer smaller than $\left( 6N\right) ^{1/3}$%
, it implies that $\mu <\epsilon _L$. One can then decrease the value of $L$
step by step, until a value of $L$ is found for which $N_L\geq N$ but $%
N_{L-1}<N$.

Fortunately, this procedure can also be used if $\mu $ is smaller than the
lowest energy of the model. It then results in a negative value of $L,$
which is related to the chemical potential in (\ref{eq:mu}) by formally
filling out $\epsilon _L$ with $L$ negative.

A fresh refining routine is then started to find $\alpha \in [0,1[$, which
is bracketed as required for sure-fire root finding programs. The actual
determination of $\alpha $ (for both cases $\mu >\frac 32\hbar w$ and $\mu <%
\frac 32\hbar w$) is straightforward from the equation for $N$ if written in
the appropriate form for numerical treatment: 
\begin{equation}
N=\sum\limits_{\nu =0}^\infty \frac 12\frac{\left( \nu +1\right) \left( \nu
+2\right) }{1+b^{L-\nu -\alpha }}=\sum\limits_{\nu =0}^\infty \frac 12\left(
\nu +1\right) \left( \nu +2\right) \left\{ 
\begin{array}{ll}
\frac 1{1+b^{L-\nu -\alpha }} & \text{if }\nu \leq L-\alpha \\ 
\frac{b^{-L+\nu +\alpha }}{b^{-L+\nu +\alpha }+1} & \text{else}
\end{array}
\right. .
\end{equation}

The determination of the chemical potential along these lines presents no
numerical difficulties. Because the Fermi energy is of order $\hbar w\left(
6N\right) ^{1/3},$ it seems natural to express $kT$ in units of $\hbar
wN^{1/3}$. This scaling factor turns out to be surprisingly good, as is
shown in Fig. 1 where the temperature dependence of the chemical potential $%
\mu \left( T\right) $ in units of $\mu \left( 0\right) $ is plotted against $%
t=\frac{kT}{\hbar wN^{1/3}}$. Introducing a density of states and making the
continuum approximation ($k_BT\ll \hbar w$) the corresponding scaling law 
\cite{Butts,Silvera} with the Fermi energy is implicit.

\subsection{The free energy}

Having determined the chemical potential, the free energy ${\Bbb F}_F\left(
N\right) =-\frac 1\beta \ln {\Bbb Z}_F\left( N\right) $ can be evaluated
from (\ref{eq:Zintegral}) 
\begin{eqnarray}
{\Bbb F}_F\left( N\right) &=&{\Bbb F}_F^{\left( 0\right) }\left( N\right) -%
\frac 1\beta \ln \left( \int_0^\pi \Psi \left( \theta \right) d\theta
\right) , \\
{\Bbb F}_F^{\left( 0\right) }\left( N\right) &=&-\frac 1\beta \ln \frac{\Xi
_F\left( u\right) }{u^N},
\end{eqnarray}
where ${\Bbb F}_F^{\left( 0\right) }\left( N\right) $ is the zero-order
steepest descent result, which would be obtained from the
``grand-canonical'' treatment. The correction factor involving $\Psi \left(
\theta \right) $ accounts for the finite number of particles. For reference, 
$\Psi \left( \theta \right) $ is shown for $N=10$ in Fig. 2 as a function of 
$\theta $ for various values of the temperature. If $N$ increases, $\Psi
\left( \theta \right) $ becomes increasingly concentrated near the origin $%
\theta =0$ for nonzero temperatures.

The resulting free energy per particle as a function of temperature is shown
in Fig. 3--5, for $N=1,$ $N=10$ and $N=100$ respectively, in units of $\hbar
w\left( 6N\right) ^{1/3}$ proportional to the Fermi energy. For comparison,
the contribution ${\Bbb F}_F^{\left( 0\right) }\left( N\right) $ is also
plotted (dashed lines). As expected, this steepest descent contribution
becomes increasingly accurate if the number of fermions increases.

\subsection{The internal energy}

The contribution of the relative degrees of freedom to the internal energy 
\begin{equation}
{\Bbb U}_F=\frac d{d\beta }\left( \beta {\Bbb F}_F\right) ={\Bbb F}_F-T\frac %
d{dT}{\Bbb F}_F
\end{equation}
can be obtained from the free energy obtained above by numerical
differentiation. The internal energy per particle is plotted in Fig. 6 in
units of $\hbar w\left( 6N\right) ^{1/3}$ proportional to the Fermi energy.
A similar scaling law is observed as for the chemical potential.

In one dimension the thermodynamical properties of harmonically interacting
bosons and fermions can be derived from each other. This case has been
studied in I. In higher dimensions, the fermion internal energy is smoothly
decreasing with decreasing temperature, whereas for the boson case it shows
sudden changes in slope, related to the condensation.

To within the numerical accuracy, the results are in agreement with the
standard description from the generating function treatment using 
\begin{equation}
\sum_{\nu =0}^\infty \epsilon _\nu n_\nu =\hbar w\sum_{\nu =0}^\infty \frac 1%
2\frac{\left( \nu +1\right) \left( \nu +2\right) \left( \nu +3/2\right) }{%
1+e^{-\beta \left( \mu -\epsilon _\nu \right) }}.
\end{equation}

These results can also be compared to the internal energy ${\Bbb U}_{F,\text{%
rec}}\left( N\right) $ which one would obtain from the recurrence relation.
In terms of the expressions for $z_N$ discussed above one then obtains 
\begin{equation}
{\Bbb U}_{F,\text{rec}}\left( N\right) =\hbar w\left( \frac b{z_N}\frac{%
\partial z_N}{\partial b}+\frac 32N+3\sum_{j=1}^N\frac{jb^j}{1-b^j}\right) .
\label{eq:Ucan}
\end{equation}
This calculation is in practice only feasible for a limited number of
particles $N\leq 10,$ and for these cases it coincides within the numerical
accuracy with the results plotted in Fig. 6.

\section{The groundstate energy}

In this section the low-temperature limit will be considered. By counting
the number of occupied energy levels, the dominant contribution to the
partition function can then easily be calculated, taking into account the
degeneracy $\frac 12\left( \nu +1\right) \left( \nu +2\right) $ of the
levels with energy $\epsilon _\nu =\hbar w\left( \nu +\frac 32\right) $. The
calculation is done first with the Fermi level $L$ fully occupied. The
number of particles $N_F$ required for this assumption to hold is 
\begin{equation}
N_F=\sum_{\nu =0}^L\frac 12\left( \nu +1\right) \left( \nu +2\right) =\frac 1%
6\left( L+3\right) \left( L+2\right) \left( L+1\right) .  \label{eq:NF}
\end{equation}
Consequently, the Fermi energy is of order $\left( 6N\right) ^{1/3}\hbar w.$
The total energy ${\Bbb U}_F$ associated to the case with the $L$-th level
fully occupied is 
\begin{equation}
{\Bbb U}_F=\sum_{\nu =0}^L\frac 12\left( \nu +1\right) \left( \nu +2\right)
\hbar w\left( \nu +\frac 32\right) =\frac 18\hbar w\left( L+3\right) \left(
L+1\right) \left( L+2\right) ^2.
\end{equation}
For a limited number of particles, the number of particles and the energy $%
{\Bbb U}_F$ are shown in Table \ref{TableFermi}.

For an arbitrary number $N$ of fermions not filling the Fermi level
completely, the determination of the ground state energy is slightly more
involved. We first determine the number of particles $N_F\leq N$ which fills
the level $L$. From (\ref{eq:NF}) it follows that the highest fully occupied
level $L$ is given by 
\begin{equation}
L=%
\mathop{\rm integer}%
\left( \left( 3N+\frac 19\sqrt{3^6N^2-3}\right) ^{1/3}+\frac 13\left( 3N+%
\frac 19\sqrt{3^6N^2-3}\right) ^{-1/3}-2\right) .
\end{equation}
From this level $L,$ one can determine the corresponding $N_F$ and ${\Bbb U}%
_F.$

The resulting formula for the leading term in the partition function for $N$
particles in the zero temperature limit is $b^{3N/2}b^{M_N},$ which defines
the power $M_N$ as 
\begin{equation}
M_N=\left( L+1\right) \left( N-\frac 1{24}\left( L+2\right) \left(
L+3\right) \left( L+4\right) \right) .
\end{equation}

The ground state energy $E_0$ with $N$ particles is $E_0={\Bbb U}_F+\left(
N-N_F\right) \hbar w\left( L+1+\frac 32\right) $ because the remainder of
the particles is in the level $L+1,$ and consequently 
\begin{equation}
E_0=-\frac 1{24}\hbar w\left( L+4\right) \left( L+3\right) \left( L+2\right)
\left( L+1\right) +N\hbar w\left( L+1+\frac 32\right) .
\end{equation}

\section{Conclusion and Discussion}

In this communication we have given a short review of the calculation
techniques for fermions, described in I and II for identical particles in
general. Next a numerical analysis of the chemical potential and of the free
energy is made for a given expectation value of the number of particles as a
function of temperature. In this analysis, we could easily illustrate what
the consequences are of the minus sign coming from the anti-symmetric
representation of the permutation group in the expression for the partition
function. Even when the expressions are known analytically, the plot of a
relatively smooth function such as the free energy requires special
techniques as a consequence of numerical instabilities due to a sign problem
in the recurrence relation for the partition function. The necessity of such
techniques can be checked by attempting a calculation of a few limits, which
lead to the application of De l'H\^{o}pital's rule many times, even
proportional to the square of the number of particles in the system.

It should be noted that the model contains only spin-polarized fermions. In
quantum dots, its production would require a magnetic field. In our model
this field is not included. However as we have shown in I, the expressions
for the partition function for fermions in the presence of an external
magnetic field can be obtained with the same calculation technique. The
influence of the magnetic field on the chemical potential and the specific
heat of our model has not been studied yet for fermions. In alkali metal
vapors the spin polarization of the fermions would be inherent to the
experimental technique \cite{Stoof}.

The chemical potential and the internal energy exhibit a scaling law in the
sense that it is an almost universal function of the temperature when
plotted in the indicated scaled units. Although we strongly suspect that the
scaling comes via the Fermi level of the confined system as is the case in
the continuum limit, we have no mathematical proof of this observation in
the low temperature case $k_BT\ll \hbar w.$

We did not compare the present approach with other theories using the same
or an analogous model. It should however be stressed that as far as we could
check we presented here the first results obtained with a new scheme for the
evaluation of path integrals for fermions.

\acknowledgments
\label{mark:acknowledgments}A discussion with W. Krauth on related topics is
gratefully acknowledged. This work is performed in the framework of the FWO
projects No. 1.5.729.94, G. 0287.95 and WO.073.94N (Wetenschappelijke
Onderzoeksgemeenschap, Scientific Research Community of the FWO on
``Low-Dimensional Systems''), the ``Interuniversity Poles of Attraction
Program -- Belgian State, Prime Minister's Office -- Federal Office for
Scientific, Technical and Cultural Affairs'', and in the framework of the
BOF NOI 1997 projects of the University of Antwerpen.

\label{mark:references}

\label{mark:tables}

\begin{table}[tbp] \centering%
\begin{tabular}{l}
$
\begin{array}{l}
z_2=b\left( 3+b^2\right) \\ 
z_3=b^2\left( 3+10b+6b^2+6b^3+7b^4+3b^5+b^7\right) \\ 
z_4=b^3\left( 1+15b+27b^2+62b^3+63b^4+87b^5+80b^6+87b^7+O\left( b^8\right)
\right) \\ 
z_5=b^5\left( 6+37b+105b^2+231b^3+413b^4+669b^5+921b^6+1197b^7+O\left(
b^8\right) \right) \\ 
z_6=b^7\left( 15+75b+290b^2+687b^3+1590b^4+2994b^5+5304b^6+8388b^7+O\left(
b^8\right) \right) \\ 
z_7=b^9\left(
20+135b+543b^2+1645b^3+4206b^4+9381b^5+19131b^6+35802b^7+O\left( b^8\right)
\right) \\ 
z_8=b^{11}\left(
15+173b+780b^2+2871b^3+8296b^4+21453b^5+49110b^6+104723b^7+O\left(
b^8\right) \right) \\ 
z_9=b^{13}\left(
6+135b+847b^2+3612b^3+12348b^4+36166b^5+93972b^6+223572b^7+O\left(
b^8\right) \right) \\ 
z_{10}=b^{15}\left(
1+57b+615b^2+3261b^3+13503b^4+45345b^5+134610b^6+357933b^7+O\left(
b^8\right) \right)
\end{array}
$%
\end{tabular}
\caption{Reduced partition function  $z_N$  for $N$ = 2, 3, ..., 10. \label{TablezN}}%
\end{table}%

\begin{table}[tbp] \centering%
\begin{tabular}{l}
$z_{10}/b^{15}= 
\begin{array}[t]{l}
1+57b+615b^2+3261b^3+13503b^4+45345b^5+134610b^6+357933b^7+879054b^8+2010684b^9+4345128b^{10}
\\ 
+8918028b^{11}+17522121b^{12}+33074766b^{13}+60269475b^{14}+106291845b^{15}+182005221b^{16}+303159450b^{17}
\\ 
+492298273b^{18}+780509769b^{19}+12101\,16969b^{20}+18367\,96808b^{21}+27328%
\,28889b^{22}+39890\,23158b^{23} \\ 
+57179\,09554b^{24}+80544\,27489b^{25}+1\,11580\,11888b^{26}+1\,52106%
\,15846b^{27}+2\,04163\,94163b^{28}+2\,69955\,46500b^{29} \\ 
+3\,51805\,18626b^{30}+4\,52045\,91771b^{31}+5\,72943\,27336b^{32}+7\,16529%
\,67164b^{33}+8\,84501\,88705b^{34}+10\,78019\,80392b^{35} \\ 
+12\,97604\,68767b^{36}+15\,42935\,02560b^{37}+18\,12786\,76665b^{38}+21%
\,04869\,19309b^{39}+24\,15851\,83659b^{40} \\ 
+27\,41281\,60656b^{41}+30\,75731\,38975b^{42}+34\,12822\,38438b^{43}+37%
\,45505\,13570b^{44}+40\,66182\,87529b^{45} \\ 
+43\,67101\,70877b^{46}+46\,40538\,23241b^{47}+48\,79244\,29070b^{48}+50%
\,76635\,37606b^{49}+52\,27216\,27332b^{50} \\ 
+53\,26698\,44520b^{51}+53\,72332\,78185b^{52}+53\,62897\,29798b^{53}+52%
\,98885\,99492b^{54}+51\,82338\,66459b^{55} \\ 
+50\,16872\,64528b^{56}+48\,07367\,47215b^{57}+45\,59865\,53010b^{58}+42%
\,81161\,40777b^{59}+39\,78628\,50832b^{60} \\ 
+36\,59783\,30982b^{61}+33\,32107\,42599b^{62}+30\,02657\,98386b^{63}+26%
\,77945\,02825b^{64}+23\,63641\,32672b^{65} \\ 
+20\,64551\,18098b^{66}+17\,84442\,48222b^{67}+15\,26113\,40019b^{68}+12%
\,91339\,21739b^{69}+10\,81016\,37948b^{70} \\ 
+8\,95194\,77403b^{71}+7\,33263\,42790b^{72}+5\,94028\,45476b^{73}+4\,75902%
\,66765b^{74}+3\,76992\,18057b^{75}+2\,95259\,67669b^{76} \\ 
+2\,28593\,93796b^{77}+1\,74930\,02580b^{78}+1\,32289\,33857b^{79}+98853%
\,84267b^{80}+72976\,28572b^{81}+53215\,14726b^{82} \\ 
+38322\,46167b^{83}+27251\,03340b^{84}+19129\,72224b^{85}+13255%
\,07922b^{86}+906297026b^{87}+611414355b^{88} \\ 
+406837455b^{89}+266998784b^{90}+\allowbreak
172746528b^{91}+110189817b^{92}+69257169b^{93}+42900069b^{94}+26170152b^{95}
\\ 
+\allowbreak
15728664b^{96}+9303894b^{97}+5421180b^{98}+3106746b^{99}+1753755b^{100}+%
\allowbreak 972819b^{101}+531755b^{102} \\ 
+285204b^{103}+150903b^{104}+78197b^{105}+40038b^{106}+\allowbreak
20010b^{107}+9928b^{108}+4758b^{109}+2298b^{110} \\ 
+1056b^{111}+492b^{112}+213b^{113}+\allowbreak
99b^{114}+39b^{115}+18b^{116}+7b^{117}+3b^{118}+b^{120}
\end{array}
$%
\end{tabular}
\caption{Reduced partition function  $z_{10}$. \label{Tablez10}}%
\end{table}%

\begin{table}[tbp] \centering%
\begin{tabular}{c}
$
\begin{tabular}{|rrr|rrr|rrr|}
\hline
$\nu _F$ & $N_F$ & $U_F/\hbar w$ & $\nu _F$ & $N_F$ & $U_F/\hbar w$ & $\nu
_F $ & $N_F$ & $U_F/\hbar w$ \\ \hline
$1$ & $4$ & $9$ & $10$ & $286$ & $2574$ & $60$ & $39711$ & $\frac{3693123}2$
\\ 
$2$ & $10$ & $30$ & $15$ & $816$ & $10404$ & $70$ & $62196$ & $3358584$ \\ 
$3$ & $20$ & $75$ & $20$ & $1771$ & $\frac{58443}2$ & $80$ & $91881$ & $%
\frac{11301363}2$ \\ 
$4$ & $35$ & $\frac{315}2$ & $25$ & $3276$ & $66339$ & $90$ & $129766$ & $%
8953854$ \\ 
$5$ & $56$ & $294$ & $30$ & $5456$ & $130944$ & $100$ & $176851$ & $\frac{%
27058203}2$ \\ 
$6$ & $84$ & $504$ & $35$ & $8436$ & $234099$ & $150$ & $585276$ & $66721464$
\\ 
$7$ & $120$ & $810$ & $40$ & $12341$ & $\frac{777483}2$ & $200$ & $1373701$
& $\frac{416231403}2$ \\ 
$8$ & $165$ & $\frac{2475}2$ & $45$ & $17296$ & $609684$ & $250$ & $2667126$
& $504086814$ \\ 
$9$ & $220$ & $1815$ & $50$ & $23426$ & $913614$ & $300$ & $4590551$ & $%
\frac{20795\,19603}2$ \\ 
$10$ & $286$ & $2574$ & $55$ & $30856$ & $1319094$ & $400$ & $10827401$ & $%
\frac{65289\,22803}2$ \\ \hline
\end{tabular}
$%
\end{tabular}
\caption{Number of particles $N_F$ and total energy $U_F$ with the energy level 
$\nu_F$ fully occupied.\label{TableFermi}}%
\end{table}%

\begin{center}
{\bf Figure captions }
\end{center}

\begin{description}
\item[Fig. 1:]  Scaled chemical potential $\frac{\mu \left( T\right) }{\mu
\left( T=0\right) }$ as a function of the scaled temperature $t=\frac{kT}{%
\hbar wN^{1/3}}$ for 10, 100, 1000 and 10000 fermions. For reference, this
quantity is also plotted for 1 particle.

\item[Fig. 2:]  The integrand $\Psi \left( \theta \right) $ of equation (\ref
{eq:Psi}) for 10 fermions as a function of $\theta $ for various values of
the temperature, expressed in units of $T_{\text{ref}}=\hbar wN^{1/3}/k.$

\item[Fig. 3:]  Scaled free energy per particle $f=\frac{{\Bbb F}_F/N}{\hbar
w\left( 6N\right) ^{1/3}}$ as a function of the scaled temperature $t=\frac{%
kT}{\hbar wN^{1/3}}$ for 1 particle. For comparison, the zero-order steepest
descent contribution is also plotted (dashed line).

\item[Fig. 4:]  Same as Fig. 3 but for 10 fermions.

\item[Fig. 5:]  Same as Fig. 3 and 4 but for 100 fermions.

\item[Fig. 6:]  Scaled internal energy per particle $u=\frac{{\Bbb U}_F/N}{%
\hbar w\left( 6N\right) ^{1/3}}$ as a function of the scaled temperature $t=%
\frac{kT}{\hbar wN^{1/3}}$ for 10, 100, 1000 and 10000 fermions. For
reference, the result for 1 particle is also plotted.
\end{description}


\begin{references}
\bibitem{BDLPRE97a}  F. Brosens, J. T. Devreese, and L. F. Lemmens, Phys.
Rev. E{\bf \ 55}, 227 (1997).

\bibitem{BDLPRE97b}  F. Brosens, J. T. Devreese, and L. F. Lemmens, Phys.
Rev. E {\bf 55}, 6795{\bf \ }(1997).

\bibitem{Stringari}  S. Stringari, Phys. Rev. Lett. {\bf 76}, 1405 (1996).

\bibitem{BDLPRA97}  F. Brosens, J. T. Devreese, and L. F. Lemmens, Phys.
Rev. A{\bf \ 55}, 2453 (1997).

\bibitem{BEC1}  M. H. Anderson, J. R. Ensher, M. R. Matthews, C. E. Wieman,
and E. A. Cornell, {\sl Science }{\bf 269, }198 (1995).

\bibitem{BEC2}  K. B. Davis, M. O. Mewes, M. R. Andrews, N. J. van Druten,
D. S. Durfee, D. M. Kurn, and W. Ketterle, Phys. Rev. Lett. {\bf 75}, 3969
(1995).

\bibitem{BEC3}  C. C. Bradlet, C. A. Sacket, J. J. Tollett, and R. G. Hulet,
Phys. Rev. Lett. {\bf 75}, 1687 (1995).

\bibitem{Grossman}  S. Grossman and M. Holthaus, Z. Naturforsch. {\bf 50a},
323; 921 (1995).

\bibitem{Grossman2}  S. Grossman and M. Holthaus, Phys. Lett. A {\bf 208},
188 (1995).

\bibitem{Ketterle}  W. Ketterle and N. J. van Druten, Phys. Rev. A {\bf 54},
656 (1996).

\bibitem{Kirsten}  K. Kirsten and D. J. Toms, Phys. Rev. A {\bf 54}, 4188
(1996).

\bibitem{Haugerud}  H. Haugerud, T. Haugset, and F. Ravndal, Phys. Lett. A 
{\bf 225}, 18 (1997).

\bibitem{CohenLee}  L. Cohen and C. Lee, J. Math. Phys. {\bf 26}, 3105
(1985).

\bibitem{Krauth}  W. Krauth, Phys. Rev. Lett. {\bf 77}, 3695 (1996).

\bibitem{Butts}  D. A. Butts and D. S. Rokhsar, Phys. Rev. A {\bf 55,} 4346
(1997).

\bibitem{Silvera}  I. F. Silvera and J. T. M. Walraven, J. Appl. Phys. {\bf %
52, }2304 (1981).

\bibitem{LBD}  L. F. Lemmens, F. Brosens, and J. T. Devreese, Phys. Lett. A 
{\bf 189}, 437 (1994).

\bibitem{BDLssc}  F. Brosens, J. T. Devreese, and L. F. Lemmens, Solid State
Commun. {\bf 96}, 137 (1995).

\bibitem{LBDPR}  L. F. Lemmens, F. Brosens, and J. T. Devreese, Phys. Rev. 
{\bf E 53}, 4467 (1996).

\bibitem{Stoof}  H. T. C. Stoof, M. Houbiers, C. A. Sackett, and R. Hulet,
Phys. Rev. Lett. {\bf 75}, 3969 (1995)
\end{references}
\end{document}